\documentclass[conference]{IEEEtran}
\IEEEoverridecommandlockouts
\usepackage{cite}
\usepackage{amsmath,amssymb,amsfonts}
\usepackage{algorithmic}
\usepackage{graphicx}
\usepackage{textcomp}
\usepackage{xcolor}
\usepackage{caption} 
\def\BibTeX{{\rm B\kern-.05em{\sc i\kern-.025em b}\kern-.08em
    T\kern-.1667em\lower.7ex\hbox{E}\kern-.125emX}}
\begin{document}

\title{Automated use case diagram generator using NLP and ML\\
}

\author{
\IEEEauthorblockN{H.Rukshan Piyumadu Dias}
\IEEEauthorblockA{\textit{Department of computing} \\
Informatics Institute of Technology\\
Colombo,Sri Lanka \\
rukshan.20210046@iit.ac.lk}
\and

\IEEEauthorblockN{C.S.L.Vidanapathirana}
\IEEEauthorblockA{\textit{Department of computing} \\
Informatics Institute of Technology\\
Colombo,Sri Lanka \\
chanuka.20210284@iit.ac.lk}
\and

\IEEEauthorblockN{Rukshala Weerasinghe}
\IEEEauthorblockA{\textit{Department of computing} \\
Informatics Institute of Technology\\
Colombo,Sri Lanka \\
 rukshala.w@iit.ac.lk }
\and

\IEEEauthorblockN{M.D. Asitha Manupiya}
\IEEEauthorblockA{\textit{Department of computing} \\
Informatics Institute of Technology\\
Colombo,Sri Lanka \\
asitha.20210358@iit.ac.lk}
\and

\IEEEauthorblockN{R.M.S.J.Bandara}
\IEEEauthorblockA{\textit{Department of computing} \\
Informatics Institute of Technology\\
Colombo,Sri Lanka \\
sandaru.20200649@iit.ac.lk}
\and

\IEEEauthorblockN{Y.P.H.W.Ranasinghe}
\IEEEauthorblockA{\textit{Department of computing} \\
Informatics Institute of Technology\\
Colombo,Sri Lanka \\
heshan.20200699@iit.ac.lk}
}

\maketitle

\begin{abstract}
This paper presents a novel approach to generate a use case diagram by analyzing the given user story using NLP and ML. Use case diagrams play a major role in the designing phase of the SDLC. This proves the fact that automating the use case diagram designing process would save a lot of time and effort. Numerous manual and semi-automated tools have been developed previously. This paper also discusses the need for use case diagrams and problems faced during designing that. This paper is an attempt to solve those issues by generating the use case diagram in a fully automatic manner.\\
\end{abstract}


\section{Introduction}
Use case diagram is a type of UML diagram that is used for representing the dynamic aspect of the system\cite{b1}. It is useful when visualizing and specifying the system in the analysis and designing phase. User stories describe the functionality of a software system in a simple way\cite{b2}. Analysis and design phase is important because the software builds on top of the blueprint created in this phase\cite{b3}. Using manual tools like Visual UML, Draw.io, Rational Rose, Smart Draw etc to draw use case diagrams can be more time consuming. In a survey conducted in 2017 it has found that 54.5\% people have encountered the problem of ‘time consuming’ when it comes to drawing use case diagrams\cite{b4}. In another survey in 2017 it has found that 40.2\% of system engineers create UML diagrams manually, 23.9\% generate UML diagram using semi-automatic methods and only 13\% generate UML diagram using automatic methods\cite{b5}.  And also when drawing use case diagrams manually the system analyst has to remember the correct notation of drawing use case diagrams. These issues show the need for automation when drawing use case diagrams.\\

This paper has proposed a novel model or method for generating use case diagrams from textual user stories, by using Natural Language Processing(NLP) and Machine Learning(ML). Under the topic of  automating UML diagrams, compared to other UML diagrams like class diagram, sequence diagram and activity diagram, there are fewer researches that have been conducted on automated use case diagram generation. This research will contribute towards filling the gap of identifying actors,use cases and relationships from the user story and presenting it as a use case diagram which follows correct UML notations.\\

The remainder of this paper is structured as follows: in section 2 it reviews the literature on the same domain and parent domain. In section 3 the proposed approach will be discussed. the proposed method’s results and their discussion will be provided under section 4. Finally the conclusion of the research will be mentioned in section 5.\\

\section{Literature Review}
Limited amount of research has been done in the domain of automated use case diagram generation. Because of that reason, the research team has decided to conduct the literature review on both the same domain(Use case diagrams) and parent domain(UML diagrams) in the year range of 2006 to 2023.\\

Sharif Ahmed et al.\cite{b6} proposed a system to identify the actors and use cases from a given text by using NLP techniques. The user story will split into sentences and Tokenize those sentences using either Stemming or Lemmatization. And then for Part-Of-Speech (POS) tagging, the Stanford CoreNLP has been used. Finally, a set of heuristic rules have been used to identify the actors and use cases. Additionally, some tools like Word Sense Disambiguation (WSD) have been used to identify words with similar senses, which helps to reduce ambiguity. Also, “Anaphora Resolution” has been solved using JavaRAP by replacing nouns with their correct noun form.\\

Zakarya Alksasbeh et al. \cite{b7} have proposed a system that uses NLP. First user has to enter the text or import the text file. Then text segmentation or split text into sentences have been used. Text tokenization, Part Of Speech tagging have been applied using the Word Net library. This paper also detects grammatical errors with probabilistic parsing and spelling errors with Dictionary lookup technique. And then the Porter stemming algorithm has been used in the stemming process to eliminate all non-word tokens. To extract knowledge, a set of predefined rules has been mentioned to identify actors, use cases and relationships. And finally, the use case diagram will be created.\\

Vemuri, Chala and Fathi \cite{b8} have proposed a system which includes both NLP and ML  models for generating UML diagrams. In this proposed model, there are four main sub models. They are as follows: Pre processor , Actor classifier, Use case classifier and Post processor. In the Pre processor, tokenizing, POS tagging, Extraction of nouns and verbs, splitting and removing tags processes are done for filtering out the two subsets of actors and use cases from input text using the NLP model. Then both classifier models are used to identify the respective actors’ or use cases’ relationships. A ML model is used for those two classifiers. Finally, post processors are used to convert the classified results into the diagram. Accuracy, recall and precision tests had been run for evaluating the proposed system.\\

Narawita and Vidanage, \cite{b3} suggest a system that uses both NLP and ML models for generating UML diagrams. In the NLP model,  Sentence Splitting,  Lexical Analysis,  Syntax Analysis, Word Chunking were done to identify specific semantic elements from the input text. Then, A ML based classifier was created to identify use cases and relationships. For this classifier three different algorithms were created using Weka. They were as follows, a multi-layered perceptron which has got very high accuracy, a logistics algorithm which is significantly lower than multi layered perceptron but with higher performance and Sequence Minimal Optimization (SMO) which is lower accuracy than logistics and highest performance. After the classification, it used a Rich Based Approach to functionalities like  remove unwanted terms in the user input text,  identify specific terms in the input text and define Weka ARFF file names to read the files.\\

\section{Proposed Method}
This section describes the proposed approach that works under 4 phases. The figure 1 below shows the pipeline of the proposed approach.

\begin{figure}[htbp]
\centerline{\includegraphics[scale=0.35]{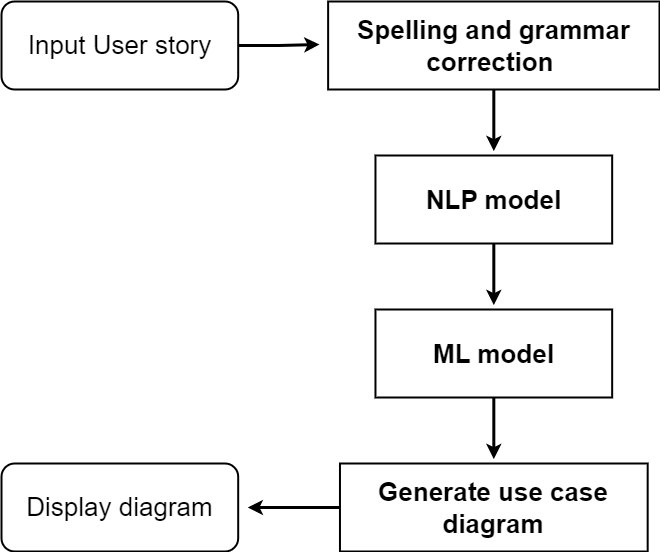}}
\caption{Proposed model pipeline}
\label{fig}
\end{figure}

\subsection{Spelling and grammar correction}\label{AA}
In this proposed approach when it comes to identifying actors and use cases from the given user story, NLP technology will be used. If the given user story contains grammatical errors it will cause inaccurate prediction. So Grammatical Error Correction or GEC is a vital task in the field of NLP\cite{b9}. Grammatical error correction is a task of identifying errors of a text and correcting it automatically\cite{b10}. A GEC model corrects errors like punctuation, spelling, tense, syntax and subject-verb agreement errors in a text\cite{b11}. \\

For this proposed approach a GEC model called, “Gingerit” that have built using gingersoftware.com API will be used. Gingerit library will detect and correct types errors in text that have been mentioned above \cite{b12}. In this phase the initial user story will be taken as input and correct the grammar errors and pass the grammatically correct user story to the NLP model.\\

\subsection{NLP approach}
Natural Language Processing (NLP) is a subcategory of Artificial intelligence and Machine learning which allows computers to understand and manipulate natural human language\cite{b13}. In this approach the NLP model will be used as the key model to identify the actors and use cases with their relationship.\\

The NLP model will receive grammatically correct user story from the GEC model. First the user story will be divided into sentences using sentence segmentation. Next using word tokenization, the sentences will be divided into words. For example, “I like to read books” after tokenization it will be, [I], [like], [to], [read], [books]. Tokenization helps to review every and each word carefully.\\

After word tokenization, Part of speech (POS) tagging will be implemented. POS provides valuable information about words or tokens \cite{b14}. POS tagging will be used to identify whether the word is a noun, verb, adjective, pronoun, conjunction, preposition, numeral, article or interjection\cite{b15}. After POS tagging, Dependency parsing will be applied to every word in the sentence. Dependency parsing is a technique used to analyze grammatical relations between words in a sentence\cite{b16}. It would create a tree or graph data structure of the sentence breaking down the words or tokens based on the grammatical relation. Using dependency parsing leads to an accurate NLP model.\\

The algorithms would identify the subject of the sentence using “nsub” dependency parser and where it is not an “PRON” (pronoun) in POS tagging. Then it would identify the object and the verb of the sentence. To identify the object, dependency parser of “dobj” and the verb will be identified using POS tagging of “VERB”. After identifying the subject, object and the verb, lemmatization will be applied to those. Lemmatization is used to convert a token or a word to its root word\cite{b17,b18}.

\begin{table}[ht]
    \centering
    \setlength{\tabcolsep}{25pt}
\renewcommand{\arraystretch}{1.5}
\resizebox{8cm}{!}{
    \begin{tabular}{|c|c|}
         \hline
\textbf{Word} & \textbf{Lemmatization} \\  \hline
 Buys & Buy \\  \hline
Better & Good \\ \hline
changing & change \\
\hline
    \end{tabular}
    }
    \caption{word and its lemmatization}
\end{table}

Finally unique subjects will be identified as actors and the combination of verb and object will be identified as the use case. The association relationship between actors and the use cases will be identified as follows. When a new actor is identified, all the use cases after that will be considered as the use cases of that actor until another new or different actor appears. The extracted data of the use case diagram will be stored in a dictionary data structure.\\

\subsection{ML approach}
The ML model will receive a dictionary data structure from the NLP model that contains actors and use cases. The ML model will be used to filter and remove unnecessary use cases from identified use cases in the NLP model. More than 700 use cases that have been extracted from the NLP model have been used in the data-set along with whether its a use case or not. To perform the classification on this model, Naive Bayes algorithm which is a supervised classification method will be used. In some cases, the Naive Bayesian classifier is an effective and simple probabilistic classification method \cite{b19}.
After filtering and removing the unnecessary use cases from the data structure, the updated data structure will be passed on to the diagram generation model.\\

\subsection{Diagram generation}
In this phase the final identified actors and use cases from ML model as a data structure of key-value pairs. Now the final identified data needs to be converted into a visual use case diagram. To visually represent use case diagram several tools have been identified,

\begin{table}[ht]
    \centering
\renewcommand{\arraystretch}{1.7}
\resizebox{9cm}{!}{
    \begin{tabular}{| c | p{1.7cm} | p{2.2cm} |}
         \hline
\textbf{Tool Name} & \textbf{Use correct UML notation} & \textbf{Support use case diagrams} \\  \hline
 plantUML & \checkmark & \checkmark \\  \hline
yUml & X & \checkmark   \\ \hline
Chart mage & \checkmark & X  \\ \hline
Zen UML & \checkmark & X   \\ \hline
Umple & \checkmark & X   \\ \hline
UML Graph & \checkmark & X   \\ \hline
Dot UML & \checkmark & X   \\ 
\hline
    \end{tabular}
    }
    \caption{Diagram generating tools comparison}
\end{table}

After a comparison on various tools, plantUml has been selected as the tool to visualize the use case diagram.
To generate the use case diagram using plantUml it is required to write pseudo code. It also can be automated using the final updated actor-use case dictionary. For example, below code would be the plantUml code to display the use case diagram for, actor - “Customer” use case - “buy product”,\\

\begin{verbatim}
@startuml
    left to right direction
    actor "Customer" as Cu
    rectangle {
      usecase "buy product" as UC1
    }
    Cu --> UC1
@enduml
\end{verbatim}

\section{Results and Discussion}
This paper proposes an approach to generate use case diagrams automatically. First GEC will be performed. The results of GEC are impressive. It fixes most of all grammatical errors but when it comes to fixing spelling issues, the GEC model struggles a bit. When a word is misspelled, the GEC model will replace that with a totally different word sometimes. 

\begin{table}[ht]
    \centering
    \setlength{\tabcolsep}{20pt}
\renewcommand{\arraystretch}{1.3}
\resizebox{8cm}{!}{
    \begin{tabular}{| c | c |}
         \hline
 User stories count & 8 \\  \hline
 Actual actors count & 15 \\  \hline
 Actual use cases count & 28 \\  \hline
Identified actors count & 13 \\ \hline
Identified use cases count & 20 \\
\hline
    \end{tabular}
    }
    \caption{Proposed model evaluation}
\end{table}

By using 8 user stories where one user story contains 20 to 35 words, 13 out of 15 actors have been identified correctly which is about 86.67\% and 20 out of 28 use cases have been identified correctly which is about 71.43\%.\\

Since the ML model is a supervised classification model, the confusion matrix will be used to evaluate the model. 

\begin{table}[ht]
    \centering
  \setlength{\tabcolsep}{10pt}
\renewcommand{\arraystretch}{1.3}
\resizebox{9cm}{!}{
    \begin{tabular}{| c | c | }
        \hline
 TRUE POSITIVE (TP) & FALSE NEGATIVE (FN) \\  \hline
FALSE POSITIVE (FP)  & TRUE NEGATIVE (TN) \\ 
\hline
    \end{tabular}
    }
    \caption{Confusion table}
\end{table}

Accuracy = (TP + TN) / TOTAL \\
Precision = TP/ (TP+FP) \\
Recall = TP/ (TP+FN) \\
F1 score =  2*(Precision* Recall) / (Precision + Recall) \\

Using the confusion matrix several metrics can be calculated. For example the accuracy, precision, recall score and F1 score. The accuracy is defined as the ratio of number of correct predictions by total number of predictions. The ML classification model of this study contains an accuracy of 75.52\%. The precision describes the quality of the positive predictions. The recall and the F1 score of the ML model is 76\% and 72\%.
After automatically identifying the actors, use cases and relationships and then generating the use case diagram, there still can be some issues. To prevent that this system will provide manual editing of the use case diagram. Where users can add and remove actors or use cases, rename actors or use cases and swap the connection between one actor and a use case to another actor.

For the below user story the generated use case diagram has been shown in figure 02. \\
\textit{“A customer calls a car repair shop to make an appointment for an oil change. The receptionist checks the availability of the mechanic and schedules the appointment for the next available time slot.”}

\begin{figure}[htbp]
\centerline{\includegraphics[scale=0.45]{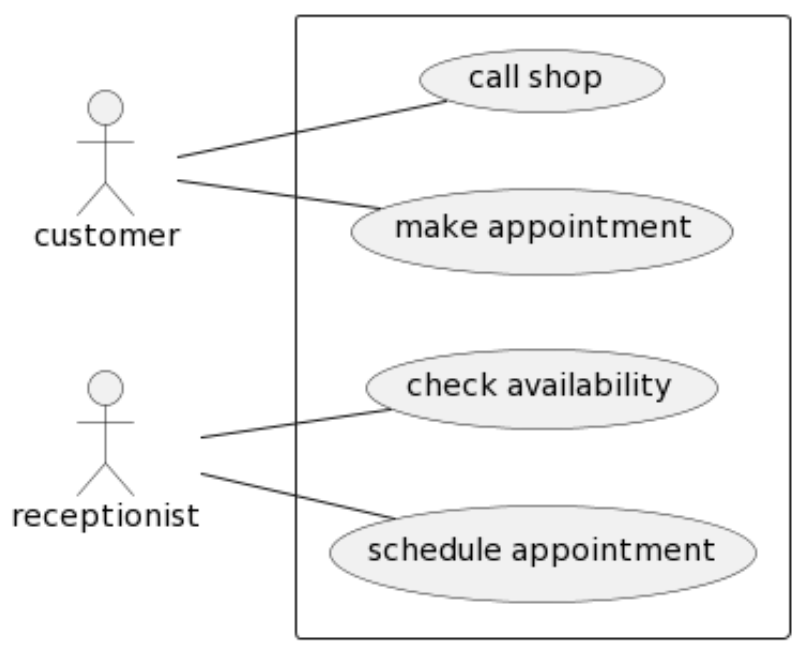}}
\caption{Generated use case diagram.}
\label{figure 02}
\end{figure}

User stories that have been used to evaluate this system have not been used to train or test the model before.

\section{Conclusion}
In conclusion, this paper proposed a method to generate the use case diagram for the given user story by using NLP and ML models. It also discussed different tools and techniques that have been considered and their advantages and disadvantages. In addition several limitations of this proposed method have been identified. This model will not identify complex relationships like include and exclude, it will only identify association relationships. Also the user has to enter the name of the system manually, it will not identify a suitable name from the given user story. Finally, use case diagrams can be generated only for English language user stories.

\section*{Acknowledgment}

We would like to thank our supervisor, Ms.Rukshala Weerasinghe for her guidance and encouragement of this work, throughout the research.

\end{document}